\documentclass[12pt]{article}

\usepackage{epsfig}
\usepackage{amssymb}
\usepackage{subfigure}
\usepackage{graphicx}
\usepackage{color}
\usepackage{cite}

\usepackage{jheppub} 

\begin{document}

\baselineskip 6mm
\renewcommand{\thefootnote}{\fnsymbol{footnote}}


\newcommand{\nc}{\newcommand}
\newcommand{\rnc}{\renewcommand}



\newcommand{\tcb}{\textcolor{blue}}
\newcommand{\tcr}{\textcolor{red}}
\newcommand{\tcg}{\textcolor{green}}


\def\ba{\begin{array}}
\def\ea{\end{array}}
\def\be{\begin{eqnarray}}
\def\ee{\end{eqnarray}}
\def\nn{\nonumber\\}


\def\ct{\cite}
\def\la{\label}
\def\eq#1{(\ref{#1})}


\def\a{\alpha}
\def\b{\beta}
\def\g{\gamma}
\def\G{\Gamma}
\def\d{\delta}
\def\D{\Delta}
\def\e{\epsilon}
\def\et{\eta}
\def\ph{\phi}
\def\Ph{\Phi}
\def\ps{\psi}
\def\Ps{\Psi}
\def\k{\kappa}
\def\l{\lambda}
\def\L{\Lambda}
\def\m{\mu}
\def\n{\nu}
\def\th{\theta}
\def\Th{\Theta}
\def\r{\rho}
\def\s{\sigma}
\def\S{\Sigma}
\def\ta{\tau}
\def\o{\omega}
\def\O{\Omega}
\def\pr{\prime}


\def\half{\frac{1}{2}}
\def\goto{\rightarrow}

\def\na{\nabla}
\def\grad{\nabla}
\def\curl{\nabla\times}
\def\div{\nabla\cdot}
\def\pa{\partial}
\def\fr{\frac}

\def\bra{\left\langle}
\def\ket{\right\rangle}
\def\lb{\left[}
\def\lc{\left\{}
\def\ls{\left(}
\def\lp{\left.}
\def\rp{\right.}
\def\rb{\right]}
\def\rc{\right\}}
\def\rs{\right)}

\def\vac#1{\mid #1 \rangle}


\def\td#1{\tilde{#1}}
\def\check{ \maltese {\bf Check!}}


\def\Tr{{\rm Tr}\,}
\def\det{{\rm det}}
\def\text#1{{\rm #1}}


\def\bc#1{\nnindent {\bf $\bullet$ #1} \\ }
\def\ch {$<Check!>$ }
\def\ss {\vspace{1.5cm}}
\def\inf{\infty}

\begin{titlepage}

\hfill\parbox{5cm} { }

 
\vspace{25mm}

\begin{center}
{\Large \bf  Time Evolution of Entanglement Entropy \\
in Holographic FLRW Cosmologies}

\vskip 1.5  cm
   {Chanyong Park$^{a}$\footnote{e-mail : cyong21@gist.ac.kr}}

\vskip 0.5cm

{\it $^a$ Department of Physics and Photon Science, Gwangju Institute of Science and Technology,  Gwangju  61005, Korea}

\end{center}

\thispagestyle{empty}

\vskip2cm


\centerline{\bf ABSTRACT} \vskip 4mm

\vspace{0.5cm}

To understand the time-dependent quantum correlation in expanding universes, we study the time-dependent entanglement entropy in the braneworld model. If we take into account a generalized string cloud geometry caused by uniformly distributed open strings, cosmologies on the braneworld result in the standard Friedmann-Lema\^{i}tre-Robertson-Walker cosmologies with various matter contents. On the dual field theory side, open strings are reinterpreted as a fundamental matter, while the black hole mass corresponds to the excitation energy of massless gauge bosons. In this work, we show how the string cloud geometry is matched to various braneworld cosmologie,s for example, eternal inflation, radiation-, and matter-dominated universes. Then, we investigate how the entanglement entropy evolves in those expanding universes.

\vspace{2cm}

\end{titlepage}





\section{Introduction}

Recently, quantum entanglement has been one of the main research areas to understand quantum features of various physical systems both in high energy physics and in condensed matter physics. Despite importance of  quantum nature, it is still a difficult problem to figure out the quantum entanglement of strongly interacting systems. In this situation, there was a very interesting and fascinating conjecture in the string theory. This is called holography or the AdS/CFT correspondence which claims that a strongly interacting quantum field theory (QFT) has a one-to-one map to a classical gravity theory defined in a one-dimensional higher anti-de Sitter (AdS) space \cite{Maldacena:1997re,Gubser:1998bc,Witten:1998qj,Witten:1998zw}. Based on the holography conjecture, in this work, we will investigate the quantum entanglement entropy and its time evolution in various expanding universes.   

In general, a two-dimensional conformal field theory (CFT) is special in that it is invariant under infinitely many conformal and modular transformations. These large symmetries allow us to determine the quantum entanglement entropy exactly even for strongly interacting systems \cite{Calabrese:2004eu,Calabrese:2009qy}. Intriguingly, it was shown that the holographic calculation in a three-dimensional AdS space, which is dual of a two-dimensional CFT, can reproduce the same results obtained in CFT \cite{Ryu:2006bv,Lewkowycz:2013nqa}. This work has been further generalized to higher-dimensional CFTs deformed by various relevant or marginal operators \cite{Ryu:2006ef,Solodukhin:2008dh,Casini:2009sr,Casini:2011kv,Myers:2012ed,Takayanagi:2012kg,Klebanov:2012yf,Nishioka:2014kpa}. In these holographic calculations, the Ryu-Takayanagi (RT) formula has been widely used  \cite{Ryu:2006bv,Lewkowycz:2013nqa}. The RT formula first assumes a dual geometry which is invariant under time translation. Then, it claims that the entanglement entropy of a deformed CFT can be represented as the area of the minimal surface extending to the dual geometry. Because of the time translation invariance, the minimal surface can be defined in a hypersurface at any given time and results in the time-independent entanglement entropy.

Now, let us take into account a time-dependent entanglement entropy. To describe it, we first break the time translation invariance \cite{Hartman:2013qma,Liu:2013iza,Liu:2013qca,Maldacena:2012xp,Liu:2012eea}. This implies that we must consider a dual geometry whose metric is time-dependent. Due to the nontrivial time dependence, we cannot directly apply the RT formula. It was argued that on the time-dependent background one must use the Hubeny-Rangamanni-Takayanagi (HRT) formula, instead of the RT formula, in which the minimal surface also extends to the time direction  \cite{Hubeny:2007xt}. Sometimes, the HRT formula is called the covariant formulation. Although the HRT formula is conceptually manifest, it is not easy to calculate the time-dependent holographic entanglement entropy exactly except for several simple cases \cite{Chu:2016pea}. Interestingly, it was argued that the RT formula even in the time-dependent geometries can provide the leading contribution to the HRT formula at a given time at least in the UV regime \cite{Koh:2018rsw}. Following this argument, the leading behavior of the time-dependent entanglement entropy has been studied in the eternally inflating universe. It has been shown that the leading time dependence of the entanglement entropy still satisfies the area law determined by the physical distance \cite{Koh:2018rsw}. However, there are still several important issues to be resolved. The first one is to calculate the higher-order corrections which may significantly modify the late-time behavior of the entanglement entropy. The second is how we can calculate the entanglement entropy in universes expanding by power-laws. Although an eternally inflating cosmology is holographically realized by an AdS space with a dS boundary, we still do not know what the exact dual geometries expanding by power-laws are.
In the standard cosmology, there are several types of cosmological universes allowing different power-law expansions. Thus, understanding the time evolution of the entanglement entropy remains as an important topic to understand the quantum nature in cosmology. 

In the present work, we take into account another holographic model that can describe a variety of expanding universes containing the expected power-law expansions. The braneworld (or Randall-Sundrum) model claims that we are living on the brane at which two bulk geometries are bordered \cite{Randall:1999ee,Chamblin:1999by,Chamblin:1999ya,Lee:2007ka,Oh:2009cz}. Originally, the first Randall-Sundrum model has been studied to explain the hierarchy issue of two very different energy scales. Later, the second Randall-Sundrum model has been improved to the braneworld model which can explain the inflationary cosmology with the graceful exit  \cite{Randall:1999vf,Kraus:1999it,Park:2000ga}. In the present work, we will show that the gravity theory involving appropriate bulk matters can describe other expanding universes with the power-law expansion. According to the AdS/CFT correspondence, those bulk matters like open strings and black hole mass can be identified with fundamental matters and radiations of the dual field theory, respectively. When we consider a four-dimensional expanding universe in which the fundamental matter or radiations are uniformly distributed, the standard cosmology shows that the scale factor increases by the power-law, for example, $\ta^{2/3}$ for the fundamental matter and $\ta^{1/2}$ for radiations where $\ta$ indicates the cosmological time. When we regard the previous bulk matters in the dual gravity theory, intriguingly, the braneworld model leads to the same power-law behaviors in the braneworld cosmology. 
   
Another remarkable point of the braneworld model is that it is possible to calculate the time-dependent entanglement entropy of various expanding universes. In the braneworld model, the expanding universe is represented as the radial motion of the brane in the bulk space. Therefore, even when we consider the static bulk spaces, the tension of the brane and the nontrivial bulk metric cause a nontrivial radial motion of the brane. Since the bulk metrics of the braneworld model are static, we can use the RT formula instead of the HRT formula. Nevertheless, the radial motion of the brane provides the time-dependent boundary conditions for the minimal surface which leads to the nontrivial time evolution of the entanglement entropy. In the eternal inflationary era, we show that the leading time dependence of the entanglement entropy is proportional to the area of the entangling surface as discussed in Ref. \cite{Koh:2018rsw}. In addition, we also calculate the higher order corrections to the time-dependent entanglement entropy. In the matter- and radiation-dominated eras representing the power-law expansion, the entanglement entropy of a small subsystem size follows the area law, while the large subsystem leads to the volume law rather than the area law. It was shown that the similar volume law of the entanglement entropy occurs in the IR regime of the RG flow and represents the thermal entropy which corresponds to the thermalization of massless gauge bosons \cite{Casini:2008cr,Blanco:2013joa,Park:2015hcz,Kim:2016jwu}. As a consequence, the entanglement entropy in the radiation-dominated era increases by $\ta$ for a small subsystem and by $\ta^{3/2}$ for a sufficiently large subsystem. In the matter-dominated era, on the other hand, the entanglement entropy increases by $\ta^{4/3}$ for a small subsystem and by $\ta^2$ for a large subsystem.

The rest of this paper is organized as follows. In Sec. \ref{Section:2}, we briefly discuss the braneworld model in which the braneworld cosmology is determined by the radial motion of the brane. In order to realize the FLRW metric on the brane, in Sec.  \ref{Section:3}, we discuss the dual gravity theory including appropriate matter fields which can represent the matter- and radiation-dominated eras. In Sec.  \ref{Section:4}, we investigate the time-dependent entanglement entropy in the expanding universes by applying the RT formula. Finally, we conclude this work with some remarks in Sec.  \ref{Section:5}.


\section{Radial motion of a brane in the braneworld model \la{Section:2}}

It would be interesting to understand the quantum correlations between two regions in the expanding universe. Even in the holographic setup, it is not an easy task to calculate the time-dependent entanglement entropy. In the expanding universe, although it was argued that the RT formula can provide a good leading approximation to the HRT formula \cite{Koh:2018rsw}, one has to use the HRT formula instead of the RT formula due to the explicit time-dependence of the metric. When applying the HRT formula, there are two difficulties in determining the entanglement entropy of the expanding universe. One is that it is not easy to construct a dual geometry which allows a Friedmann-Lema\^{i}tre-Robertson-Walker (FLRW) metric of the boundary spacetime. The other is that, even when the dual geometry is known, it is not easy to evaluate the entanglement entropy by applying the HRT formula. In the present work, we take into account another model called the braneworld model which has two merits to understand the time-dependent entanglement entropy in the expanding universe. The first one is that the FLRW metrics for various expanding universes can be easily realized. The second thing is that, in spite of the fact that the resulting entanglement entropy is time-dependent, we can still use the RT formula instead of the HRT formula. Due to these reasons, it is possible to calculate the time-dependent entanglement entropy in the braneworld model. To see more details, in this section, we briefly discuss the braneworld model. 

Let us first assume that ${\cal M}_\pm$ are two five-dimensional bulk spaces with each own well-defined metric, $g^{(\pm)}_{MN}$, and that they are bordered through a four-dimensional brane $\pa {\cal M}$. Then the induced metrics on both sides of the brane must be reduced to the same metric to get a unique metric on the brane. This requirement was called the first Israel's junction condition \cite{Israel:1966rt}, which fixes the tangential components of two bulk metrics The second junction condition treats the derivatives of the bulk metrics in the radial direction perpendicular to the brane. Although the normal components of the metrics are continuous at the border, their derivatives are generally not due to a non-vanishing stress tensor of the brane. If we further require the first normal derivatives of the bulk metrics to be continuous at the border, this constraint leads to the second Israel junction equation, When the tension of the brane is given, a radial motion of the brane is governed by this second junction equation. In the braneworld model  \cite{Randall:1999ee,Randall:1999vf,Chamblin:1999ya}, the radial motion of the brane is directly associated with the cosmology on the brane. In order to investigate possible cosmologies appearing in the braneworld model, we discuss the details of the braneworld model by using the holographic renormalization technique.  
 
Let us consider the following five-dimensional gravity action
\be
S = S_{{\cal M} _+} + S_{ {\cal M}_-} + S_{\pa {\cal M}} ,
\ee
where $S_{{\cal M} _\pm} $ indicate two gravity actions determining the bulk metrics $g^{(\pm)}_{MN}$ and the remaining $S_{\pa {\cal M}}$ denotes the action of the brane. More precisely, the gravity action $ S_{{\cal M} _\pm}$ defined on ${\cal M}_\pm$ is given by
\be
S_{\pm} = \fr{1}{2 \k^2} \int_{{\cal M}_{\pm}} d^{5} x \sqrt{- g} \ \ls {\cal R} - 2 \L^{({\pm})} + {\cal L}^{({\pm})}_m  \fr{}{}\rs - \fr{1}{\k^2} \int_{\pa {\cal M} } d^4 x \sqrt{- \g} K^{({\pm})} ,
\ee
where the last term is the Gibbons-Hawking term which is needed to obtain a well-defined Einstein equation. Here, $\g_{\m\n}$ is the induced metric on the brane and ${\cal L}^{({\pm})}_m$ denotes the Lagrangian of bulk matter fields.  From now on, we assume a negative cosmological constant
\be
\L^{({\pm})} = - \fr{6}{R_{\pm}^2} ,
\ee
where $R_{\pm}$ is the AdS radii for ${\cal M} _\pm$. In general, the action $S_{\pm}$ can have different cosmological constants and matter contents.  In the present work, for simplicity, we assume that two bulk geometries have the same cosmological constant and matter contents. If one consider different cosmological constants and matter contents, one can find a variety of different cosmologies on the brane.

If we require translational and rotational symmetries on the brane, the corresponding bulk metric has the following form 
\be		   \la{metric:GeneralAn}
ds^2 = g_{MN} dx^M dx^N =  - A(r ) d t ^2  + B(r ) dr^2 + C(r)\ \d_{ij} dx^i dx^j ,
\ee
where $i,j = 1, \cdots, 3$. Before discussing the junction equation, it must be noted that the radial ranges of ${\cal M}_{\pm}$ is restricted to a finite or semi-infinite region due to the existence of the brane. Denoting the radial position of the brane as $\bar{r}$, the range of the radial coordinate in the bulk spaces is limited to $0 \le r \le \bar{r}$ or $\bar{r} \le r\le \infty$. However, if we further require the reflection invariance under $r \to 2 \bar{r} - r$, the radial range of two bulk spaces reduces to $0 \le r \le \bar{r}$ and $\bar{r} \le r \le 2 \bar{r}$. This $Z_2$ symmetry was used to construct the first braneworld model \cite{Randall:1999ee}. Although the braneworld model technique can be applied to the general case without the $Z_2$ symmetry, we hereafter focus on the case with the $Z_2$ symmetry because the later case is sufficient to present important features we are interested in. Anyway, if the geometry of ${\cal M}_-$ is determined in the present setup, the geometry of ${\cal M}_+$ is automatically fixed due to the $Z_2$ symmetry.

After rewriting the bulk metric $g_{MN}$ as the form of the ADM decomposition, 
\be
g_{MN} dx^M dx^N  = B (r) \ dr^2 + \g_{\m\n} dx^\m dx^\n , 
\ee
we can derive an on-shell gravity action by applying the equation of motion. Due to the equation of motion, the on-shell gravity action usually reduces to a boundary term on the brane. The variation of the boundary action with respect to the boundary metric is given by \cite{Chamblin:1999ya}
\be
\delta S_{\pm}= \int_{\pa {\cal M}} d^4 x \sqrt{-\g} \ \pi^{({\pm})}_{\mu\nu}  \delta {\g}^{\mu\nu},
\ee
with the canonical momentum of the metric
\be			\la{result:BoundaryStressM}
\pi^{({\pm})}_{\mu\nu}  = \frac{1}{\sqrt{\g}} \frac{\d S_{\pm}}{\delta \g^{\mu\nu}}  = - \frac{1}{2 \k^2}  \left( K^{(\pm)}_{\mu\nu} - \g_{\mu\nu} K^{(\pm)} \right) ,
\ee
where $K_{\mu\nu} = \nabla_\m n_\n$ indicates an extrinsic curvature tensor at the boundary. In the holographic renormalisation procedure, this canonical momentum corresponds to the dual stress tensor defined in the boundary spacetime. Due to the $Z_2$ symmetry we imposed, the extrinsic curvatures of two bulk spaces satisfy $K^{(+)}_{\mu\nu} = - K^{(-)}_{\mu\nu}$ where the minus sign naturally appears due to the opposite direction of two unit normal vectors in ${\cal M}_\pm$. Defining $K_{\mu\nu}  = K^{(-)}_{\mu\nu}$, the resulting boundary stress tensors become
\be			\la{result:BoundaryStressP}
\pi^{(\pm)}_{\mu\nu}  =  \pm \frac{1}{2 \k^2}  \left( K_{\mu\nu} - \g_{\mu\nu} K \right) .
\ee

Note that the above boundary stress tensors are defined at the same position. Nevertheless, two boundary stress tensors derived in ${\cal M}^{(\pm)}$ are not same. Why does this discrepancy occur? The reason is that the brane also has its own non-vanishing stress tensor. Consequently, the brane's stress tensor must cancel the discrepancy of two boundary stress tensors mentioned before. Denoting the brane's stress tensor as $T_{\m\n}$, it must satisfy 
\be
\pi^{(+)}_{\mu\nu}  - \pi^{(-)}_{\mu\nu}  = T_{\m\n} ,
\ee 
which is the second Israel junction condition. Assuming that the brane is in a ground state with a constant energy density and pressure, the brane's action is simply given by
\be
S_{\pa {\cal M}} = - \fr{2 \s}{\k^2} \int_{\pa {\cal M}} d^4 x \ \sqrt{-\g} ,
\ee
where $2 \s/\k^2$ corresponds to a brane's tension and $\pa {\cal M}$ indicates the brane's worldvolume. In this case, the brane's stress tensor is given by 
\be
 T_{\m\n}  = \fr{1}{\sqrt{-\g}} \fr{\pa S_B}{\pa \g^{\m\n}} =  \fr{ \s}{\k^2} \g_{\m\n} .
\ee
Applying the $Z_2$ symmetry discussed before, the Israel junction equation finally reduces to the following simple form
\be
K_{\m\n} = - \fr{\s}{d-1} \g_{\m\n} .
\ee

From the holographic viewpoint, the brane plays a role of the boundary for two bulk spaces in which the dual field theory resides. In this case, the position of the brane is identified with the energy scale of the dual field theory. If we are interested in physics at a certain fixed energy scale, it is natural to take into account a non-dynamical brane (or boundary) lying at a finite radial distance. In the braneworld model, however, we are interested in the cosmology on the brane which must be time-dependent. Since the cosmology on the brane is associated with the radial motion of the brane, it is better to take the radial position of the brane as a function of time, $\bar{r}(t)$. Under this parameterization, we rewrite the junction condition as a more explicit form in terms of metric components.
When the brane is moving in the radial direction, a unit normal vectors on the brane is given by
\be
n_M &=& \fr{\sqrt{A B}}{\sqrt{A-B \dot{r}^2}} \lc \dot{r}, -1,0, 0 ,0 \rc  ,
\ee
where $r$ means a position of the brane. In terms of the normal vector, the extrinsic curvature tensor is defined by $K_{MN} = \g^P_M \g^Q_N \nabla_P n_Q$ with $\g_{MN} = G_{MN} - n_M n_N$. The spatial components of the extrinsic curvature tensor result in (see Ref. \cite{Chamblin:1999ya} for more details)
\be
K_{ij} = - \fr{\sqrt{AB}}{A} \fr{C'}{C} \fr{1}{\sqrt{A-B \dot{r}^2}} \g_{ij} ,
\ee
where the prime indicates a derivative with respect to $r$. As a consequence, the junction equation reduces to
\be
\fr{C'}{C} = \fr{\s}{d-1} \fr{\sqrt{AB}}{A}  \sqrt{A-B \dot{r}^2} .
\ee
When the bulk metric is given, the junction equation determines the radial motion of the brane with a velocity $\dot{r}$.

For reinterpreting the brane's radial motion as the cosmology on the brane, we must introduce a cosmological time $\ta$ defined on the brane 
\be
- d \ta^2 =  - A dt^2 + B dr^2 .
\ee
After we replace $C(r)$ by $a(r)^2$ and regard $r$ as a function of the cosmological time $\ta$ instead of the bulk time $t$, the induced metric on the brane finally becomes
\be			\la{metric:FLRW}
ds_\S^2 = - d \ta^2 + a(\ta)^2 \ \d_{ij} dx^i dx^j  .
\ee
This is nothing but the FLRW metric representing the time evolution of the universe. In terms of the cosmological time, the junction condition is rewritten as
\be		\la{result:RadialMotion}
\ls \fr{dr}{d\ta} \rs^2= \fr{\s^2}{3^2}  \fr{C^2}{C'^2} - \fr{1}{B}.
\ee
If the bulk metric is known, the corresponding cosmology on the brane is uniquely determined by the junction equation.

\section{Cosmology on the brane \la{Section:3}}

In the previous section, we studied the general formula representing the relation between the brane's motion in the bulk space and the time evolution in the braneworld. In this section, we try to construct an appropriate gravity theory which can describe the cosmology of the braneworld with various matter contents like a vacuum energy, non-relativistic matters and radiations.

\subsection{Standard cosmology in four-dimensional flat space}

Before studying the cosmology in the braneworld model, for later comparison we briefly summarize the standard cosmology of a four-dimensional expanding universe \cite{Kinney:2009vz}. The cosmology of a four-dimensional expanding universe can be described by the FLRW metric in \eq{metric:FLRW} where $a(\ta)$ is called the scale factor. The scale factor represents how rapidly the universe expands. Due to the existence of the nontrivial scale factor in the FLRW metric, it is worth noting that a distance $|d\vec{x}|$ defined in the comoving frame is not physical. Instead, the physical distance is given by  $a(\ta)  | d\vec{x}|$.

In the standard cosmology, the time-dependence of the scale factor is governed by the Friedmann equation
\be
\ls \fr{\dot{a}}{a} \rs^2 = \fr{\k^2}{3} \r   ,
\ee
and the continuity equation
\be
0=\dot{\r} + 3 \ls \fr{\dot{a}}{a} \rs (\r + p) ,
\ee
where $\r$ and $p$ are an energy density and pressure of matters contained in the universe. Assuming that the matter is an ideal gas satisfying $ p = w \r$, the value of the equation of state parameter $w$ characterizes what kind of the matter is contained in the universe. For example, when $w=-1$ the matter corresponds to the vacuum energy or cosmological constant. If $w=0$, the matter is called dust which corresponds to non-relativistic particles with zero pressure. For $w=1/3$, lastly, the matter reduces to radiations corresponding to relativistic massless fields like gauge bosons. 

Solving the continuity equation with the equation of state parameter of an ideal gas, the energy density can be rewritten in terms of the scale factor
\be
\r \sim a^{-3(1+w)} .
\ee
Plugging this relation to the Friedmann equation, the scale factor except for $w = -1$ is determined as a function of the cosmological time
\be
a (t) \sim \ta^{\fr{2}{3(1+w)}}. 
\ee
This result shows that the scale factor in the radiation-dominated era ($w=1/3$) increases by $a \sim \ta^{1/2}$ as time elapses, while it increases by $a\sim \ta^{2/3}$ in the matter-dominated era ($w=0$). For $w=-1$, the scale factor increases exponentially with time, $a \sim e^{H \ta}$, and represents eternal inflation.

\subsection{Bulk geometry for the braneworld model}

In the previous section, we briefly discussed the possible cosmological solutions relying on the matter distributed in the universe. In the braneworld model, how can we realize these cosmologies including various different kinds of the matter? In this section, we try to construct a specific five-dimensional  gravity theory which may allow us to rederive the known results of the standard cosmology in the braneworld model.

Let us first consider a five-dimensional gravity theory without a bulk matter field. Then, due to the negative cosmological constant, the most general geometric solution is given by an AdS black hole (or black brane) metric
\be
ds^2 =  \fr{r^2}{R^2} \ls -   f(r) d t ^2 +  \d_{ij} dx^i dx^j  \fr{}{} \rs   +\fr{R^2}{r^2 f(r)} dr^2 ,
\ee
with a blackening factor
\be		\la{solution:AdSBH}
f(r) = 1- \fr{m}{r^4}  ,
\ee
where $m$ indicates a black hole mass. According to the AdS/CFT correspondence, a gravity theory having an asymptotic AdS geometry is dual to a conformal SU(N) gauge theory at a UV fixed point. Especially, an AdS black hole solution corresponds to a thermal system of such a gauge theory due to the well-defined temperature. It has been shown that the boundary stress tensor of the AdS black hole, after an appropriate holographic renormalisation procedure, is proportional to $N^2 m$, where $N$ is the rank of the gauge group. In this case, the $N^2$ dependence indicates that the matter content of the dual field theory follows an adjoint representation under the gauge group transformation. Therefore, we can identify the black hole mass with the excitation energy of the massless gauge bosons. To clarify this identification further, there exists another important remarkable point. For a $(d+1)$-dimensional AdS black hole,  it was known that the energy density and pressure derived from the boundary stress tensor satisfy the relation $p = \r/(d-1)$. For $d=4$, this relation shows that the matter of the dual boundary field theory is a relativistic massless field with the equation of state parameter $w=1/3$. This is another evidence for the previous identification between the black hole mass and the excitation energy of the gauge bosons in the dual field theory.

Now, let us think of bulk matter fields representing fundamental matters of the dual field theory. From the field theory point of view, a fundamental matter is a field transformed by a fundamental representation of the gauge group. In the string theory, it was well known that one end of an open string follows a fundamental representation due to the Chan-Paton factor. Therefore, the fundamental matter on the brane can be realized by many open strings whose one end is attached to the brane. For more details, we consider open strings on an AdS spacetime. Since the open strings are one-dimensional objects, the gravity action containing uniformly distributed $N$ open strings is written as \cite{Chakrabortty:2011sp,Chakrabortty:2016xcb}
\be
S = \fr{1}{2 \k^2} \int d^{5} x \sqrt{-g} \ls {\cal R} - 2 \L \rs  -\fr{3 {\cal T}}{4} \sum_{i=1}^N  \int d^2 \xi_i \sqrt{-h} h^{\a\b} \pa_{\a} x^{M} \pa_{\b} x^{N} g_{MN}  ,
\ee
where $\fr{3}{4} {\cal T}$ is a tension of open strings and $h_{\a\b}$ is an induced metric on the string. Here, the factor $\fr{3}{4}$ was introduced for later convenience. From this action, the Einstein equation including the gravitational backreaction of open strings reads
\be			\la{equaton:EinsteinStringC}
R_{MN} - \half R g_{MN} + \L g_{MN} = \k^2 T_{MN} ,
\ee 
with 
\be
T^{MN} = - \fr{3 {\cal E}}{2} \ \fr{\sqrt{-h}}{\sqrt{-g}}  \ h^{\a\b} \pa_{\a} x^{M} \pa_{\b} x^{N} ,
\ee
where ${\cal E} = N {\cal T}/V$ is an energy density of open strings with an appropriately regularized three-dimensional volume $V$ perpendicular to $\xi^\a$. In order to represent open strings whose one ends are attached to the brane, we take into account open strings extended to the radial direction. Then, such a string configuration can be well expressed in the static gauge with $\xi^0=t$ and $\xi^1=r$. The solution of this gravity theory was known as the string cloud geometry and studied in \cite{Letelier:1979ej,Stachel:1980zr,Stachel:1980zs,Gibbons:2000hf,Herscovich:2010vr,Chakrabortty:2011sp,Chakrabortty:2016xcb}.

Intriguingly, the gravity theory with uniformly distributed open strings allows a simple and analytic solution satisfying  weak and dominant energy conditions \cite{Chakrabortty:2011sp,Chakrabortty:2016xcb}. The metric of the string cloud geometry is given by
\begin{equation}
ds^2 = \frac{r^2}{R^2} \left(- f(r) dt^2  + \d_{ij} dx^i dx^j\right)  + \frac{R^2}{r^2 f(r)} dr^2,
\end{equation}
with the nontrivial metric factor 
\begin{equation}		\la{res:blackholefactor}
f(z) = 1  -  \fr{{\cal E}}{r^{3} } . 
\end{equation}
which resembles a black hole solution due to the existence of a horizon. To make a black hole geometry, a well-localized matter is usually required at the centre of the black hole. If we consider uniformly distributed either particles or open strings in the flat spacetime, we cannot expect the existence of a black hole-like geometry. However, this is not true for the AdS case. An AdS space has a nontrivial warping factor which makes a three-dimensional spatial volume perpendicular to the radial direction approach zero at the centre of the AdS space. This fact implies that, even when open strings are uniformly distributed,  its energy density becomes high at the centre due to the warping factor of the AdS space. This is the reason why the black hole-like geometry appears in the AdS space with uniformly distributed open strings.

Comparing the string cloud geometry with the five-dimensional AdS black hole solution in \eq{solution:AdSBH}, we easily see that the blackening factors of two-black hole solutions show different power behaviors. For the black hole solution, the matter must be well localized in the radial direction.  However, the string cloud geometry was constructed by one-dimensional objects extended to the radial direction. Due to the different dimensions of two objects in the radial direction, the blackening factor of the string cloud geometry has a different power from that of the black hole. The existence of the horizon in the string cloud geometry allows us to define temperature. The horizon is located at
\be
r_h = {\cal E}^{1/3} ,
\ee
and it leads to temperature
\be
T_H =  \fr{3 {\cal E}^{1/3}  }{4 \pi R^2} .
\ee
Although temperature is well defined,  temperature is not an essential concept in the string cloud geometry. The reason is that, unlike the black hole case, temperature of the string cloud geometry is not a free parameter because it is perfectly fixed by the energy density of open strings. Again, we would like to emphasize that the string cloud geometry is the dual of the gauge theory containing the fundamental matter due to the Chan-Paton factor of the open strings. 

We can further consider a generalization of the string cloud geometry. If we put an additional zero-dimensional matter into the centre of the string cloud geometry, the string cloud geometry can allow a generalized geometric solution 
\begin{equation}			\la{metric:GStringCloud}
ds^2 = \frac{r^2}{R^2} \left(- f(r) dt^2  + \d_{ij} dx^i dx^j\right)  + \frac{R^2}{r^2 f(r)} dr^2,
\end{equation}
with the following blackening factor
\be
f(z) = 1  -  \fr{{\cal E}}{r^{3}} -     \fr{m}{r^{4}} ,
\ee
which is still a solution of the Einstein equation in \eq{equaton:EinsteinStringC} and corresponds to the combination of the ordinary AdS black hole and the string cloud geometry. Hereafter, we call this a generalized string cloud geometry, for convenience. It is worth emphasizing that ${\cal E}$ and $m$ in the generalized string cloud geometry are related to the energy density of the fundamental matter and gauge bosons, respectively. Since we are interested in the cosmologies of the matter-dominated and radiation-dominated eras, the generalized string cloud geometry \eq{metric:GStringCloud} is one of the good candidates to represent holographically such cosmologies in the braneworld model.

In the generalized string cloud geometry in \eq{metric:GStringCloud}, the radial motion of the brane is determined by \eq{result:RadialMotion}
\be		
\ls \fr{dr}{d\ta} \rs^2=  \ls \fr{\s^2}{36}  - \fr{1}{R^2} \rs r^2  + \fr{\r}{R^2}  \fr{1}{r} +   \fr{m}{R^2} \fr{1}{r^2} ,
\ee
and the induced metric on the brane becomes
\be
ds_\S^2 = - d \ta^2 +  \fr{r(\ta)^2}{R^2}  \ \d_{ij} dx^i dx^j  .
\ee
As shown in these relations, the radial position of the brane, $r$, is directly related to the scale factor of the braneworld, $a(\ta) = r(\ta)/R$. In order to know what kind of cosmology appears in the braneworld, we consider several specific parameter regions which reproduce the same results of the standard cosmology.

\begin{enumerate}

\item Time-independent universe

For an AdS space with ${\cal E}=m=0$, the motion of the brane is determined only by the tension of the brane and the curvature radius of the AdS space. If the tension of the brane has a critical value given by  
\be
\s_c = \fr{6}{R} ,
\ee
the brane does not move in the radial direction and the scale factor of the braneworld becomes time-independent.

\item Eternal inflationary era

If the brane in the AdS space has a tension different from the critical value $\s_c$, its radial motion is governed by
\be			\la{result:BraneVel}
 \fr{dr}{d\ta}  =   \sqrt{ \fr{\s^2}{36}  - \fr{1}{R^2} }  \ r   .
\ee
Then, the radial position of the brane in terms of the cosmological time is determined as
\be			\la{result:ScaleFactorInf}
r (\ta )  = r_i \ e^{H \ta} ,
\ee
with a Hubble constant
\be
H = \fr{\sqrt{\s^2 - \s_c^2}}{6} ,
\ee
where $r_i$ is the position of the brane at $\ta=0$, which must be determined by an appropriate initial condition. The braneworld model for $\s \ne \s_c$ is equivalent to the standard cosmology with $w=-1$. Since the scale factor in \eq{result:ScaleFactorInf} shows an eternal acceleration, from now on, we call it an eternal inflation.

\item Matter-dominated era 

Recalling that open strings in the bulk corresponds to fundamental matter on the brane, the cosmology caused by the fundamental matter on the brane can be characterized by taking $\s=\s_c$ and $m=0$. In this case, only the nontrivial contribution comes from the energy density of the open strings and the radial motion of the brane is determined by
\be
\fr{dr}{d\ta} =   \fr{\sqrt{{\cal E}}}{R}  \fr{1}{r^{1/2}}   ,
\ee
Since the radial position of the brane is the same as the scale factor of the brane cosmology, solving the above differential equation leads to the scale factor proportional to $\ta^{2/3}$
\be			\la{Result:rpositionmatter}
r (\ta ) =  \ls \fr{3}{2} \rs^{2/3} \fr{{\cal E}^{1/3}}{R^{2/3}} \ \ta^{2/3} + r_i,
\ee
where $r_i$ again indicates the initial position of the brane at $\ta=0$. In the late time era or when $r_i = 0$, the behavior of the scale factor is exactly the form of the standard cosmology with the matter having $w=0$, as expected.

\item Radiation-dominated era

Finally, let us focus on the radiation-dominated era. In this case, the radiation means a variety of massless gauge bosons whose equation of state parameter is given by $w=1/3$. The radiation dominance, as mentioned before, can be represented as the AdS black hole on the dual gravity side. As a result, the generalized string cloud geometry with $\s=\s_c$ and ${\cal E}=0$ is dual to the radiation-dominated era of the brane cosmology. In this case, the radial motion of the brane is governed by
\be
 \fr{dr}{d\ta}  =  \fr{\sqrt{m}}{R } \fr{1}{r } ,
\ee
This differential equation leads to the scale factor proportional to $\ta^{1/2}$, which is the expected form of the scale factor in the radiation-dominated era. More precisely, the resulting scale factor reads
\be
r (\ta ) =  \fr{\sqrt{2} m^{1/4}}{\sqrt{R}} \ \ta^{1/2}  + r_i.
\ee
In the late time era, this is the same as the result of the previous standard model.

\end{enumerate}

\section{Entanglement entropy in the expanding universe  \la{Section:4}}

In the previous section, we discussed the possible cosmological solutions in the braneworld and showed that the braneworld model can holographically realize the standard cosmology well. In this section, we  investigate the holographic entanglement entropy in the expanding universe. Recently, it has been shown that if the field theory and its dual geometry have a time-translation symmetry, its entanglement entropy can be easily calculated in the dual gravity theory by applying the Ryu-Takayanagi (RT) formula \cite{Ryu:2006bv,Ryu:2006ef}. However, if we are interested in the entanglement entropy in the expanding universe where the time-translational symmetry is completely broken, we must utilize the Hubeny-Rangamani-Takayanagi (HRT) formula \cite{Hubeny:2007xt}, instead of the RT formula. Despite this fact,  in the braneworld model it is still possible to use the RT formula to investigate the time dependence of the entanglement entropy in the expanding universe. The reason is that, even when the bulk geometry is given by a time-independent form, the radial motion of the brane governed by the junction equation causes an expansion of the braneworld. In this case, we can also expect the nontrivial time dependence of the entanglement entropy due to the time-dependent background spacetime on the brane \cite{Chu:2016pea}. Therefore, it would be interesting to study how the entanglement entropy evolves with time in a variety of the expanding universes.

\subsection{Entanglement entropy in a four-dimensional flat space}

To obtain more intuitions about the time evolution of the entanglement entropy in the expanding universe, let us first consider a static brane with $\s=\s_c$ and ${\cal E}=m=0$. In this case, the bulk geometry is simply given by an AdS space and the brane does not move in the radial direction. Moreover, the induced metric on the brane is just a time-independent flat Minkowski metric. Although this induced metric does not describe the expanding universe, the holographic evaluation of the entanglement entropy in this setup is helpful to understand the entanglement entropy in various expanding universes.

For convenience, we introduce a new coordinate $z=R^2/r$. Then, a five-dimensional AdS metric  in the Poincare patch can be written as
\be
ds^2 
     &=& \fr{R^2}{z^2} \ls dz^2 - dt^2 + d u^2 + u^2 d \O_{2}^2 \rs 
\ee
where $\O_{2}$ indicates a solid angle of a two-dimensional unit sphere. In the braneworld model, the range of the radial coordinate is restricted to $\bar{z} \le z \le \infty$ where $\bar{z}$ indicates the position of the brane. In this case, the brane plays a role of a finite UV cutoff from the viewpoint of the holographic renormalisation. Thus, this setup is exactly the same as the cutoff AdS space studied recently in the $T \bar{T}$-deformation \cite{Smirnov:2016lqw,Cavaglia:2016oda,McGough:2016lol,Park:2018snf}. Applying the RT formula to the cutoff AdS geometry, the entanglement entropy is associated with the area of a minimal surface extended to the cutoff AdS space. To calculate the holographic entanglement entropy, we divide a three-dimensional space into two parts, $0 \le u \le l$ and $l \le u < \infty$, with a two-dimensional sphere with a radius $l$. In this case, the two-dimensional sphere dividing the system into two parts is usually called an entangling surface. Then, the entanglement entropy between the inside and outside of the entangling surface is governed by
\be			\la{action:EntanglementE}
S_E = \fr{R^{d-1}  \O_{2}}{4 G} \int_0^l d u \ \fr{u^{2} \sqrt{1 + z'^2}}{z^{3}}  ,
\ee
where the prime means a derivative with respect to $x$. This action leads to the equation of motion, which determines the configuration of the minimal surface,
\be
0 =  1+ z'^2 + z z'' .
\ee
The general solution of this equation is given by
\be
z (x) = \sqrt{c_1 - \ls c_2 + u \rs^2 } ,
\ee
where $c_1$ and $c_2$ are two integral constants.

In order to determine the exact configuration of the minimal surface, we need to fix two undetermined integral constants by imposing appropriate two boundary conditions. To do so, first, it is worth noting that the smoothness of the minimal surface leads to $z'=0$ at $u=0$ due to the rotational symmetry. If we denote the value of $z(0)$ as $z_0$, $z_0$ corresponds to a turning point or tip of the minimal surface. In this case, the turning point gives rise to an upper bound for the range of $z$ extended by the minimal surface. Another important thing we should notice is that the properties of the turning point fix one of the undetermined integral constant to be $c_2 = 0$. When we calculate the area of the minimal surface, second, the end of the minimal surface must be identified with the entangling surface defined at the boundary. This implies that we must impose another boundary condition, $\bar{z} = z \ls l \rs$. This additional boundary condition fixes the remaining integral constant to be
\be
c_1 = l^2 + \bar{z}^2 .
\ee
Using these integral constants fixed by two natural boundary conditions, the coordinates of the minimal surface satisfies the following circular trajectory
\be
z^2 + x^2 = l^2 + \bar{z}^2 ,
\ee
where the ranges of $z$ and $x$ are restricted to $\bar{z} \le z \le z_0$ and $0 \le u \le l$ respectively. As a result, the solution satisfying all natural boundary conditions is given by
\be			
z (u) = \sqrt{l^2+ \bar{z}^2 - u^2}  .
\ee

Using the obtained solution, the resulting entanglement entropy on the brane at $\bar{z}$ becomes
\be      \la{result:RHEE}
S_E = \fr{R^{3}}{12 G } \frac{  l^{3}  \Omega_{2}  }{\bar{z}^{2} \sqrt{l^2+\bar{z}^2}}  \, _2F_1\left(\frac{1}{2},1;\frac{5}{2};\frac{l^2}{l^2+\bar{z}^2}\right)  .
\ee
In the $\bar{z} \to 0$ limit, the leading term of the entanglement entropy reduces to
\be
S_E = \fr{R^{3}}{8 G } \frac{  l^{2}  \Omega_{2}  }{\bar{z}^{2}}  + \cdots ,
\ee
where the ellipsis indicates higher order corrections. This is nothing but the well-known holographic entanglement entropy of a four-dimensional CFT \cite{Calabrese:2009qy,Ryu:2006bv,Ryu:2006ef}. In this case, the position of the brane $\bar{z}$ plays a role of an appropriate UV cutoff and $l^{2}  \Omega_{2}$ corresponds to the area of the entangling surface.


\subsection{Entanglement entropy in an eternally inflating universe}

Now, let us consider the entanglement entropy in the eternal inflationary cosmology. In the braneworld model, the eternal inflation on the brane appears when we consider an AdS bulk geometry (${\cal E}=m=0$) with a noncritical brane tension $\s \ne \s_c$. Even in this case, since the bulk geometry has nothing to do with the brane tension, the dual geometry is still given by an AdS space. The difference from the previous case is that the brane moves in the radial direction with the velocity in \eq{result:BraneVel}. Therefore, the holographic calculation of the entanglement entropy is almost the same as the previous static case. However, there exists one big difference caused by the radial motion of the brane. When we take into account the circular trajectory of the minimal surface, the boundary condition imposed on the brane must be slightly modified because the brane is moving. Requiring the end of the minimal surface to attach to the moving brane, the consistent solution must be given by a function of $\ta$ and $u$ 
\be			\la{Solution:AdS}
z (\ta,u) = \sqrt{l^2+ \bar{z}(\ta)^2 - u^2} .
\ee
It is worth to noting that the time dependence of the holographic entanglement entropy in the braneworld model appears due to the time-dependent boundary condition at $\bar{z} (\ta)$. Performing the integration in \eq{action:EntanglementE} with the obtained time-dependent solution, the resulting entanglement entropy again yields \eq{result:RHEE} with the time-dependent brane position $\bar{z}(\ta)$, instead of a constant $\bar{z}$, because \eq{action:EntanglementE} contains only the integration over $u$ which is independent of $\ta$.

In spite of the fact that the same form of the entanglement entropy obtained in the static brane again appears even in the moving brane, the physical implications of the entanglement entropy in the braneworld model dramatically changes due to the nontrivial time-dependence. The subsystem size denoted by $l$ is the size measured in the comoving frame. Due to the nontrivial scale factor on the moving brane, the size measured in the comoving frame is not a physical one. Instead, the physical size of the subsystem $L$ is given by
\be
L (\ta)= \fr{R}{\bar{z} (\ta) } \, l ,
\ee
where $\ta$ is not the conformal time but the cosmological time. Even when the subsystem has a time-independent finite size $l$ in the comoving frame, the physical size $L$ in the expanding universe changes  because the background spacetime expands with time. In general, since the scale factor $\sim 1/\bar{z} (\ta)$ relies on the matter distributed in the expanding universe, the resulting entanglement entropy and its time dependence also crucially depend on the   matter on the brane.

Regarding the moving brane whose radial position is given by a function of $\ta$, the resulting entanglement entropy is given by
\be				\la{result:TimeDHEE}	
S_E = \fr{R^{3}}{12 G } \frac{  l^{3}  \Omega_{2}  }{\bar{z}(\ta)^{2} \sqrt{l^2+\bar{z}(\ta)^2}}  \, _2F_1\left(\frac{1}{2},1;\frac{5}{2};\frac{l^2}{l^2+\bar{z}(\ta)^2}\right)  .
\ee
From the viewpoint of an observer living on the brane, the position of the brane must be reinterpreted as the scale factor relying on the cosmological time
\be
\bar{z} (\ta) = \fr{R^2}{\bar{r}(\ta)} = \fr{R^2}{r_0} \ e^{- H \ta} .
\ee
Then, the entanglement entropy in the eternal inflationary era finally results in
\be
S_E^{(inf)}=\fr{ r_0^2}{12 G R} \frac{  l^{2}  \Omega_{2}  e^{2 H \ta} }{ \sqrt{1+ \ls R^4 /r_0^2  l^2 \rs e^{- 2 H \ta}}}  \, _2F_1\left(\frac{1}{2},1;\frac{5}{2};\frac{1}{1+ \ls R^4 /r_0^2  l^2 \rs e^{- 2 H \ta}}\right)   .
\ee

Remembering that the physical distance increases with time by $L\sim l e^{H \ta}$ in the inflationary era, the above result shows that the entanglement entropy is proportional to the area of the entangling surface measured by the physical distance, $A = \pi L^2 \O_2$. In the early time of the inflation era ($H \ta \ll 1$), the small time perturbation leads to 
\be
S_E^{(inf)} &=& \frac{l r_0 \Omega _2 \sqrt{l^2 r_0^2+R^4}}{8 G R}-\frac{R^3 \Omega _2 }{16 G} \log \left(\fr{\sqrt{R^4 + l^2 r_0^2} + l r_0}{\sqrt{R^4 + l^2 r_0^2} - l r_0 }\right) \nn
 && + \fr{H l^3 r_0^3\O_2 \ta}{4 G R \sqrt{R^4 + l^2 r_0^2}}  + {\cal O} \ls  \tau^2 \rs ,
\ee
which shows that the entanglement entropy in the early time increases with time linearly. However, the entanglement entropy grows exponentially in the late time of the inflationary era  ($ H \ta  \gg 1$) 
\be
S_E^{(inf)} =\frac{l^2 r_0^2 \Omega _2 e^{2 H \tau }}{8 G R}
+\frac{R^4 \Omega _2  \ls  1 + 2  \log (2 R^2 / l r_0)  \rs}{16 G R}  -\frac{H R^3  \Omega _2 \tau }{8 G} + {\cal O} \ls  e^{- 2 H \tau }\rs  .
\ee
Here the leading term is exactly proportional to the physical area of the entangling surface, so that the area law of the entanglement entropy is still satisfied even in the expanding universe. In addition, the late time behavior, $S_E^{(inf)} \sim e^{2 H \ta}$, in the braneworld inflation is consistent with the result expected in a different  holographic model \cite{Koh:2018rsw}.

\subsection{Entanglement entropy in the radiation- and matter-dominated eras}

Now, let us take into account the entanglement entropy in the radiation- and matter-dominated eras. In the generalized string cloud geometry in \eq{metric:GStringCloud}, we already showed that the radiation- and matter-dominated eras can occurs as the cosmological solution of the braneworld. In this background, the entanglement entropy in the $z$-coordinate system is given by
\be			\la{Action:HEE}
S_E = \fr{R^{3}  \O_{2}}{4 G} \int_0^l d u\ \fr{u^{2} \sqrt{f + z'^2}}{z^{3} \sqrt{f}}  
\ee
where 
\be
f(z) = 1  - \td{{\cal E}} z^3  -\td{m} z^4  \quad {\rm with} \ \  \td{{\cal E}} \equiv \fr{{\cal E}}{R^6} , \ {\rm and} \ \td{m} \equiv \fr{m}{R^8} .
\ee
From now on, we set $R=1$ for convenience. Due to the nontrivial factor $f$ in the entanglement entropy formula, it is not  easy to find an analytic solution satisfying the equation of motion. Thus, we consider specific parameter regions in which we can get some information about the time evolution of the entanglement entropy of the expanding universe. To do so, let us remind an important feature of the minimal surface in the AdS black hole geometries. The minimal surface has to be smooth in the entire bulk geometry. This fact indicates that there exists a turning point where $z' = 0$. This turning point appears at $u=0$ due to the rotational symmetry. Denoting the turning point as $z_0 = z(0)$, it gives rise to a upper bound of $z$. In the braneworld model, on the other hand, the position of the brane $\bar{z}$ provides a lower bound of $z$. These two bounds restrict the range of $z$ extended by the minimal surface to $\bar{z} \le z \le z_0$. If the subsystem size $l$ in the comoving frame becomes smaller, $z_0$ approaches $\bar{z}$. On the other hand, when $l$ increases, $z_0$ also increases.

Now, we consider a very small subsystem size with $\bar{z} \ll \td{{\cal E}}^{-1/3}$ and $\bar{z} \ll  \td{m}^{-1/4}$. In this parameter region, the minimal surface extends only near the brane and the general string cloud geometry is slightly deviated from an AdS space. Therefore, it is possible to investigate the time evolution of the entanglement entropy perturbatively. 

\subsubsection{Radiation-dominated era}

When the brane has the critical tension $\s=\s_c$ and the background geometry is given by the AdS black hole with ${\cal E}=0$, the cosmology on the brane in the late time era corresponds to the radiation-dominated era, as mentioned before.

Now, let us investigate how the entanglement entropy in the radiation dominate era relies on the cosmological time. For convenience, we set $\td{m}=1/z_h^4$. In the small subsystem size limit $z_0 \ll z_h$, then, the configuration of the minimal surface configuration can be determined by solving the equation of motion perturbatively. To do so, we assume the perturbative form of $z(u)$ as
\be
z (u)  = z_0 (u) + \td{m}  \ z_1 (u)  +  {\cal O} \ls \fr{1}{z_h^{8}}\rs .
\ee  
In this case, $z_0$ describes the minimal surface lying in the pure AdS and is given by\eq{Solution:AdS}. To determine $z_1$ exactly, we must impose two boundary conditions. Since we assume that the end of the minimal surface is located at $\bar{z}$, $z_1$ vanishes at $u=l$ because $z_0$ already satisfies $z_0 (l) = \bar{z}$. The other boundary condition we must impose is $z'(0) = 0$ due to the smoothness of the minimal surface at $u=0$. The solution $z_1 (u)$ satisfying these two boundary conditions is given by
\be
z_1 (u) =  -\frac{\left(l^2-u^2\right) \left(2  l^4 +5 \bar{z}^4 + 6 l^2 \bar{z}^2 - 3 l^2 u^2  -4 \bar{z}^2 u^2+u^4\right)}{10 \sqrt{l^2 + \bar{z}^2-u^2}}
\ee
After substituting these solution into the entanglement entropy formula in \eq{Action:HEE}, performing the integration leads to
\be
S_E &=& \frac{c  l  \Omega _2 \sqrt{l^2 + \bar{z}^2}}{12 \bar{z}^2}
+\frac{c   \Omega _2}{24}   \log  \ls \fr{\bar{z}^2}{2 l  \left(\sqrt{l^2 + \bar{z}^2}+l \right) } \rs  \nn
&& +\frac{c \td{m}   l^3 \Omega _2}{60} \  \fr{  2 \bar{z}^2 l^2 \left(l^2-2 \bar{z}^2\right)+l^4 \left(6 \bar{z}^2-4 l^2\right) + l^2 \left(5 \bar{z}^4-8 \bar{z}^2 l^2+2 l^4\right)+2 l^6 }{ \bar{z}^4 \sqrt{l^2 + \bar{z}^2}} .
\ee
Rewriting the scale factor in the $z$-coordinate, the brane position becomes a function of the time
\be
\bar{z} = \fr{ z_h z_i  }{ z_h  + \sqrt{2} z_i  \sqrt{\ta} } ,
\ee
In the early time era ($z_i \sqrt{\ta} \ll z_h \sqrt{R}$), the entanglement entropy has the following perturbative form
\be
S_E &=&  \frac{c   l  \Omega _2 \sqrt{l^2 + z_i^2}}{12 z_i^2}
+\frac{c   \Omega _2}{24}   \log  \lb \fr{z_i^2}{2 l  \left(\sqrt{l^2 + z_i^2}+l \right) } \rb   
+\frac{c  l^5 \Omega _2 }{60} \frac{ \td{m}}{\sqrt{l^2+z_i^2}} \nn
&& + \frac{c    \Omega _2}{60 \sqrt{2} z_h } \  \fr{  2 l^5 \left(\td{m} z_i^4+10\right)+25
   l^3 z_i^2-5 l^2 z_i^2 \sqrt{l^2+z_i^2}-5 z_i^4 \sqrt{l^2+z_i^2}+5 l
   z_i^4 }{z_i \left(l^2+z_i^2\right)^{3/2}} \sqrt{\ta} \nn
   && + {\cal O} \ls \ta \rs .
\ee
This result shows that the entanglement entropy initially increases by $\sqrt{t}$. In the late time era ($\ta \to \infty$), $\bar{z}$ approaches $0$ and the entanglement entropy leads to
\be
S_E &=& \frac{c l^2  \Omega _2}{6 z_h^2} \ta 
+\frac{c l^2  \Omega _2}{3 \sqrt{2} z_h z_i}  \sqrt{\ta} 
-\frac{c  \Omega _2}{24}   \log   \ta \nn
&&   +\frac{1}{120} c  \Omega _2 \lb 2 l^4 \td{m} +\frac{10 l^2 }{z_i^2} +10  \log \left( \fr{z_h  }{2 \sqrt{2} l}  \right)+5  \rb + {\cal O} \ls \fr{1}{\sqrt{\ta}}\rs .
\ee 
In the late time of the radiation-dominated era, the entanglement entropy increases by $\ta$.

\subsubsection{Matter-dominated era}

Similar to the previous radiation-dominated era, we can also investigate how the entanglement entropy depends on the time in the matter-dominated era. To take into account the matter-dominated era, we focus on the string cloud geometry with $\s=\s_c$ and $m=0$. In a small subsystem size limit $z_0^3 \td{{\cal E}}  \ll 1$, the minimal surface configuration can be described by the following perturbative form
\be
z (u) = z_0 (u) + \td{{\cal E}}  z_1 (u) + {\cal O} \ls \td{{\cal E}}^2  \rs .
\ee
Varying the entanglement entropy formula in \eq{Action:HEE}, the equations governing $z_0 (u)$ and $z_1 (u)$ are given by
\be
0 &=& z_0''  + \frac{2 z_0'^3}{u}+\frac{3 z_0'^2}{z_0 }+\frac{2 z_0' }{u}+\frac{3}{z_0 }  , \nn
0 &=& z_1''  + \left(\frac{6 z_0'}{z_0}+\frac{6  z_0'^2+2}{u}\right) z_1'  -\frac{3   \left(z_0'^2+1\right)}{z_0^2} z_1 
   + \frac{2 z_0^3 z_0'^3}{u}+\frac{3}{2} z_0^2 \left(z_0^2-2\right) .
\ee
The solution of the first differential equation is again given by \eq{Solution:AdS}. After substituting the solution of $z_0$ into the second differential equation, we can also find an exact and analytic solution of $z_1$ but it has a very complicated form. 

Hereafter, we just focus on two specific limits corresponding to the early and late time era of the matter-dominated universe. Rewriting the brane position in \eq{Result:rpositionmatter} in terms of the $z$-coordinate, the radial position is determined to be
\be
\bar{z}=\frac{z_i}{ 1 + (3/2)^{2/3} z_i \, \td{{\cal E}}^{1/3}  \, \ta^{2/3} }  .
\ee
Assuming that the initial position of the brane $z_i$ is much smaller than $\td{{\cal E}}^{1/3} $, then the entanglement entropy is perturbatively expanded to the following form in the early time era  
\be
S_E&=& \frac{c \Omega _2}{12}  \log  \fr{\bar{z}}{ 2 l}  +\fr{c \Omega _2}{24} + \frac{c \td{\cal E} l^5 \Omega _2}{60 \bar{z}^2}+\frac{c l^2 \Omega _2}{12
   \bar{z}^2}-\frac{\bar{z}^2 c \Omega _2}{48 l^2} \nn
&=& \fr{c \Omega _2 }{12}  \log \fr{z_i}{2 l}   +\frac{c \Omega _2 \left(4 \td{\cal E} l^7+20 l^4+10 l^2 {z_i}^2-5 {z_i}^4
 \right)}{240 l^2 {z_i}^2} \nn
&& +\frac{\sqrt[3]{2 \td{\cal E}} c  \Omega _2 \lb 4 \times 3^{2/3}  \ls   \td{\cal E} l^3+ 5 \,  \rs   l^4   {z_i}  +  5 \times 3^{2/3}  \ls  {z_i}^2 -2  l^2  \rs  {z_i}^3  \rb}{240 l^2 {z_i}^2}  \ta^{2/3} + {\cal O} \ls \ta^{4/3}\rs .
\ee
This result shows that the entanglement entropy of the matter-dominated universe increases by $\ta^{2/3}$ in the early time era. Repeating the similar calculation in the late time era, the entanglement entropy has the following expansion form
\be
S_E &=& \frac{c \Omega _2  l^2}{12 \bar{z}^2}  + \frac{c \Omega _2}{12}  \log  \fr{\bar{z}}{ 2 l}  +\fr{c \Omega _2}{24}  + \frac{7 \pi  c  \Omega _2}{256} \td{\cal E} l^3 \nn
&=& \ls \frac{3}{2} \rs^{1/3} \, \frac{c \Omega _2   l^2 }{8}  \td{\cal E}^{2/3}  \ta^{4/3}
+\frac{c \Omega _2  l^2  {\td{\cal E}}^{1/3}}{ 2^{5/3} \times {3}^{1/3} {z_i}} \ta^{2/3} \nn
&& - \fr{c \Omega _2 }{18} \log \ta  +\frac{c \Omega _2  }{24} + \frac{c \Omega _2  l^2 }{12 z_i^2}  - \frac{c \Omega _2  \log \left( 18 \td{\cal E}  l^3 \right) }{ 36 }  +  \frac{7 \pi  c \Omega _2 \td{\cal E} l^3 }{256} +{\cal O} \ls \ta^{-2/3}\rs .
\ee
This resulting form indicates that the entanglement entropy increases by $\ta^{4/3}$ in the late time era of the matter-dominated universe. Comparing with that of the radiation-dominated universe, the entanglement entropy increases more rapidly in the matter-dominated universe.


\section{Discussion \la{Section:5}}

In this work, we have investigated the time-dependent entanglement entropy in various expanding universes of the braneworld model. To describe eternal inflation, two different holographic models are possible. The first one is to consider an AdS space whose boundary is given by a dS space. In this case, since the bulk metric depends explicitly on time, we must use the HRT formula instead of the RT formula. Although the HRT formula is well defined conceptually, it is not easy to calculate the holographic entanglement entropy because it is described by nontrivially coupled differential equations. If we are just interested in the leading qualitative behavior of the time evolution of the entanglement entropy, it was shown that the leading behavior of the HRT formula can be described by the RT formula. To understand the quantum entanglement of our universe, we need to know how to calculate the entanglement entropy of the other expanding universes by a power-law. The holographic model mentioned above is not applicable to describe such universes with a power-law expansion. Due to this reason, in this work, we investigated another holographic model called the braneworld model, which can easily realize the universes expanded by the power-law. In the braneworld model, the expansion on the braneworld is determined by the radial motion of the brane in the dual geometry. Moreover, since the bulk geometries are given by static ones, the RT formula instead of the HRT formula is still applicable. For the eternal inflation, we showed that the time-dependent entanglement entropy of the braneworld gives rise to the same leading behavior as the one obtained in the AdS space with the dS boundary.

To realize the radiation- and matter-dominated eras of the cosmology, we have taken into account the generalized string cloud geometry. The generalized string cloud geometry contains uniformly distributed strings and another matter localized at the centre of the background AdS space. We showed that the dual gravity theory including strings and localized matter can be identified with the braneworld containing the fundamental and adjoint matters. More precisely, the uniformly distributed strings correspond to the fundamental matter because the end of an open string transforms as the fundamental representation under the gauge group. On the other hand, the matter field localized at the centre of the AdS space leads to a Schwarzschild-type AdS black hole which is associated with the excitation energy of boundary gauge bosons following the adjoint representation. In the braneworld model, we showed that those bulk matters corresponding to the boundary fundamental and adjoint matters can realize the matter- and radiation-dominated eras of the braneworld cosmology. Intriguingly, we also showed that the braneworld model involving the fundamental and adjoint matters can reproduce the cosmological behaviors with the expected power-law expansion in the matter- and radiation-dominated eras. 

Based on these aspects of the braneworld model, we have further investigated the time-dependent entanglement entropy in various expanding universes. To describe the time-dependence of the entanglement entropy, we introduced the cosmological time instead of the conformal time. Then, the entanglement entropy of the expanding universe increases with time. For the eternally inflating four-dimensional universe the entanglement entropy increases by $S_E \sim \ta $ in the early time era, whereas it increases by $S_E \sim e^{2 H \ta}$ in the late-time era, as expected in Ref. \cite{Koh:2018rsw}. This late-time behavior is related to the fact that the leading entanglement entropy is proportional to the area of the entangling surface proposed by the RT formula. In the four-dimensional case, the area of the entangling surface is proportional to $L^2$ where $L$ is the physical distance. Since the physical distance is related to the cosmological time by $L \sim e^{H \ta}$, the late time behavior in the entanglement entropy in the eternally inflating universe is nothing but the area law of the entanglement entropy.  

In the radiation- and matter-dominated eras, the time-dependent entanglement entropy shows similar features. In the late time era of the radiation- and matter-dominated universes, the entanglement entropy also follows the area law similar to the eternal inflation case. In the radiation- and matter-dominated eras, the physical distance in terms of the cosmological time increases by $L \sim \ta^{1/2}$ and $\sim \ta^{2/3}$, respectively. As a result, the area law implies that the entanglement entropy increases by $S_E \sim \ta$ in the radiation-dominated universe and by $S_E \sim \ta^{4/3}$ in the matter-dominated universe.

\vspace{0.5cm}


\begin{acknowledgments}
C. Park was supported by Mid-career Researcher Program through the National Research Foundation of Korea grant No. NRF-2019R1A2C1006639.
\end{acknowledgments}

\vspace{0.5cm}




\bibliographystyle{apsrev4-1}
\bibliography{References}

\end{document}